\documentclass{article}

\usepackage[final]{neurips_2019}

\usepackage[utf8]{inputenc} 
\usepackage[T1]{fontenc}    
\usepackage{hyperref}       
\usepackage{url}            
\usepackage{booktabs}       
\usepackage{amsfonts}       
\usepackage{nicefrac}       
\usepackage{microtype}      
\usepackage{amsmath}
\usepackage{graphicx}
\usepackage[colorinlistoftodos]{todonotes}
\usepackage{caption}

\title{Correlation of Auroral Dynamics and GNSS Scintillation with an Autoencoder}

\author{%
	Kara Lamb\thanks{Equal Contributions} \\
	Cooperative Institute for Research in the Environmental Sciences/NOAA\\
	Boulder, CO, USA\\
	\And 
	Garima Malhotra*  \\
	University of Michigan\\
	Ann Arbor, Mi, USA\\
	\And
	Athanasios Vlontzos* \\
	Imperial College London \\
	London, UK\\
	\And
	Edward Wagstaff*\\
	University of Oxford \\
	Oxford, UK\\
	\And 
	Atılım Günes Baydin \\ University of Oxford\\ Oxford, UK
	\And 
	Anahita Bhiwandiwalla \\ Intel AI Lab \\ Santa Clara , CA, USA
	\And 
	Yarin Gal \\ University of Oxford \\ Oxford, UK
	\And
	Alfredo Kalaitzis \\ Element AI \\ London, UK
	\And
	Anthony Reina \\ Intel AIPG \\ San Diego, CA, USA\\
	\And
	Asti Bhatt \\ SRI International \\ Menlo Park, CA, USA
}

\begin{document}
	
	\maketitle
	
	\begin{abstract}
		High energy particles originating from solar activity travel along the the Earth's magnetic field and interact with the atmosphere around the higher latitudes. These interactions often manifest as aurora in the form of visible light in the Earth's ionosphere.  These interactions also result in irregularities in the electron density, which cause disruptions in the amplitude and phase of the radio signals from the Global Navigation Satellite Systems (GNSS), known as 'scintillation'. In this paper we use a multi-scale residual autoencoder (Res-AE) to show the correlation between specific dynamic structures of the aurora and the magnitude of the GNSS phase scintillations ($\sigma_{\phi}$). Auroral images are encoded in a lower dimensional feature space using the Res-AE, which in turn are clustered with t-SNE and UMAP. Both methods produce similar clusters, and specific clusters demonstrate greater correlations with observed phase scintillations. Our results suggest that specific dynamic structures of auroras are highly correlated with GNSS phase scintillations. 
	\end{abstract}
	
	\section{Introduction}
	Space weather includes phenomena associated with high energy particles originating from the sun and their expansion throughout the solar system. As the high energy particles interact and are captured by the Earth's magnetic field they are deposited in high latitudes around the two poles. Their increased concentration creates perturbations in the Earth's ionosphere which affect the propagation of radio signal communications depending on the scale of perturbations and the frequency of the signals. 
	
	The global navigation satellite system (GNSS) is a constellation of satellites around the Earth used for accurate navigation, timing and positioning. Each GNSS receiver is connected to four or more satellites receiving radio signals used to calculate accurate location and timing coordinates of the receiver. The frequency of these signals is in GHz and therefore, they interact with small scale irregularities in the ionosphere. This can cause the phase and the amplitude of the radio waves to change, resulting in degraded performance or even complete loss of lock. The changes in amplitude and phase are known as \emph{scintillations} -- this paper is focused on \emph{phase} scintillations ($\sigma_{\phi}$), as these are considered to be more important in the high latitudes \cite{spogli2009}. 
	
	At high latitudes, the interaction of high energy solar wind particles with terrestrial neutral gases in the Earth's ionosphere leads to visually observed effects such as the Aurora Borealis/Australis. Aurorae are highly dynamic phenomena that have long been thought to correlate with amplitude and phase scintillations due to similar geophysical drivers causing these phenomena. Previous research has demonstrated that the occurrence of significant phase scintillations is correlated with the presence of the aurora \citep[]{aarons2000,spogli2009,van2015}, and that specific structures in the aurora may be more correlated with high $\sigma_{\phi}$ values (e.g. \citep[]{jin2014,oksavik2015,van2015,jin2016}).
	
	Variations in the visible aurora are manifestations of variations in the geophysical drivers. Auroral images have long been classified in a few distinct classes pointing to their geophysical driving mechanisms. Several recent studies have applied deep learning methods to aurora image classification \cite[]{clausen2018, qyang2019,xyang2019}. These studies have used supervised learning approaches, which can be subjective, as they rely on the judgement of human experts to label auroral events. Human annotation can bias labels to identifying only the most prominent of features in the images, and also it relies on pre-defining specific classes that may exclude some of the salient features associated with physical processes occurring in the ionosphere. In addition to being time-consuming, labelling typically focuses on identifying application-specific features. In this work, we investigate unsupervised approaches to clustering aurora images using a general-purpose approach. Images of the Aurora Borealis originating from the THEMIS network of Northern Canada \cite[]{mende2009} are passed through a Residual Autoencoder (Res-AE) \cite{he2016,Vlontzos2019}, creating high-dimensional embeddings. The embeddings are then dimensionally-reduced and clustered with t-SNE and UMAP \cite{maaten2008, mcinnes2018}.
	We investigate the correlations of clusters to both known aurora classes and the magnitude of the phase scintillations measured at co-located GNSS receivers.
	
	\section{Proposed Method}
	\noindent\textbf{Dataset} Our dataset contains 35,277 images from THEMIS (Time History of Events and Macroscale Interactions during Substorms) All-Sky Imager (ASI) data from the Fort Simpson (FSIM) site in Northern Canada. The images are part of the THEMIS auroral dataset, taken $\pm3$ hours from midnight and represent two months of data from January and February 2015. An additional 7700 manually annotated auroral images, classified among 6 classes (\emph{arc, diffuse, discrete, moon, clear, clouds}) depending on the auroral structure \cite{clausen2018} from March 2015, are used to evaluate embeddings in the low-dimensional feature space. We note that images are classified in the moon category only if the presence of the moon makes the aurora not visible (per the discretion of the expert annotating the images). We also use measurements from a PolarRxS GNSS receiver (Septentrio 2015, Leuven, Belgium) at Fort Simpson from the Canadian High Arctic Ionospheric Network (CHAIN) to provide measurements of the scintillation indices at the same time as the aurora images.
	
\begin{figure}
		\includegraphics[width=1\textwidth]{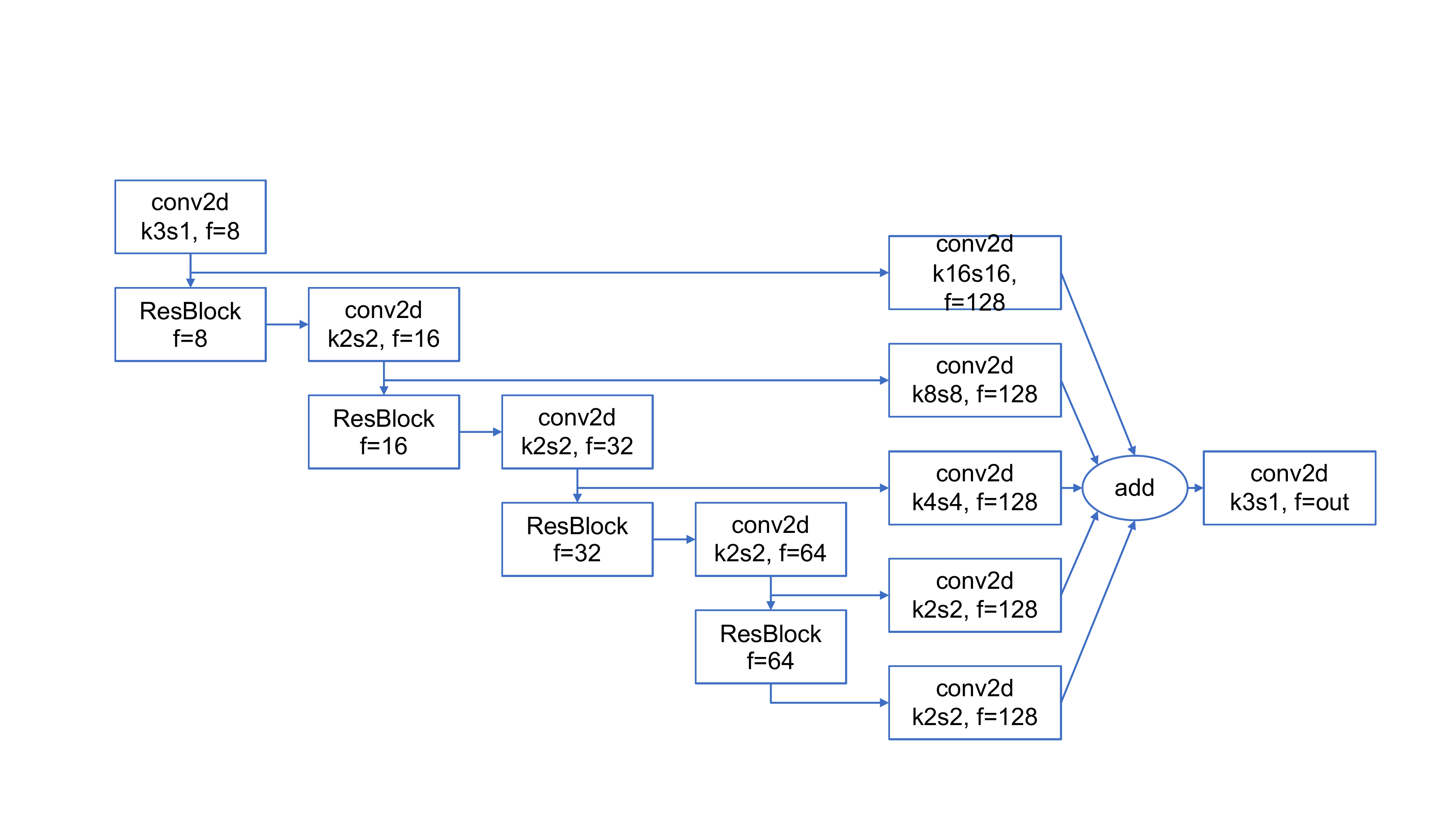}
		\caption{Res-AE Encoder \label{resnet}}
	\end{figure}

\begin{figure}
		\centering
		\begin{tabular}[b]{c}
			\includegraphics[width=\linewidth]{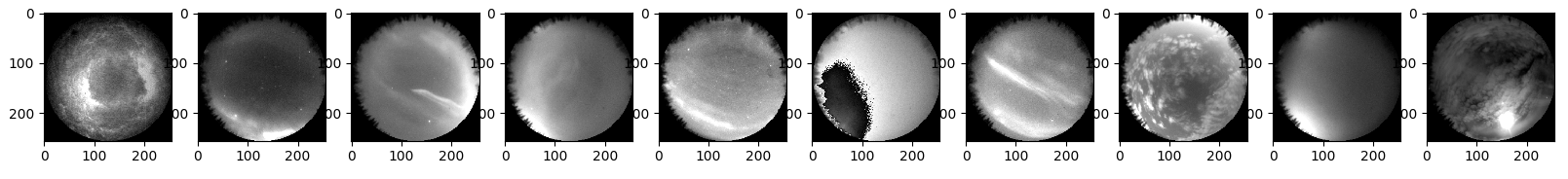} \\
			\includegraphics[width=\linewidth]{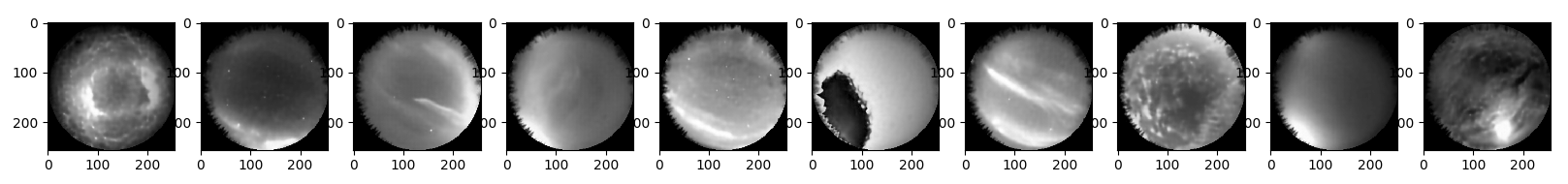} \\
		\end{tabular} \qquad
		\caption{Original (top) and reconstructed (bottom) Themis images. \label{reconstructions}}
	\end{figure}

	\noindent\textbf{Residual Autoencoder}
	In this work we use a Residual Autoencoder (Res-AE) as introduced in \cite{Vlontzos2019}.
	Figure \ref{resnet} illustrates the encoder, and the decoder mirrors the encoder.
	The overall architecture has a U-Net like structure, where sequential convolutions create multi-scale representations of the input \cite{Ronneberger2015}. Each resolution level includes a residual block \cite{he2016} and a convolutional layer with a stride of 2. The resulting feature map is passed to the next resolution level. Furthermore, a series of convolutional layers map the features from all resolution levels to an encoding of size $L_1 \times L_2 \times 128$, adding them all together to create an intermediate representation $L'$.
	A final convolutional layer then maps the intermediate representation to the encoding $L$ of size $L_1\times L_2\times L_3$.
	The size of the representation L is a critical hyperparameter as it defines the expressiveness of the encoder. In our work we found the optimal trade-off between feature space compression and expressiveness to be $32\times32\times3$.
	The multi-scale and residual nature of the autoencoder compresses the high-dimensional input images into a feature representation that accounts for multi-scale features and dependencies while filtering out some of the noise.
	Figure \ref{reconstructions} shows an example of the test and reconstructed images using the Res-AE trained on images from the FSIM Themis site. 
	
	\noindent\textbf{Unsupervised clustering of embedded latent representations}
	We investigate Barnes-Hut t-distributed Stochastic Neighbor Embedding (t-SNE) and Uniform Manifold Approximation and Projection (UMAP) for dimensionality reduction of the latent space \cite{maaten2008, mcinnes2018}. Both approaches provide lower dimensional representations of higher dimensional data, such that essential topological structures of the higher dimensional space are preserved in the density of embeddings in the lower dimensional space. In our approach, THEMIS images are mapped to the latent space $L$ using the trained Res-AE and then clustered with t-SNE or UMAP to embed the encoded latent representation into two dimensions for easier visualisation and interpretation.
	In this case we constrain the local neighborhood from which UMAP learns the manifold structure of the data to 15 nearest neighbors, as this provides a reasonable separation between the clusters.
	
	\section{Results}
\begin{figure}
		\centering
		\begin{tabular}[b]{c}
			\includegraphics[width=.42\linewidth]{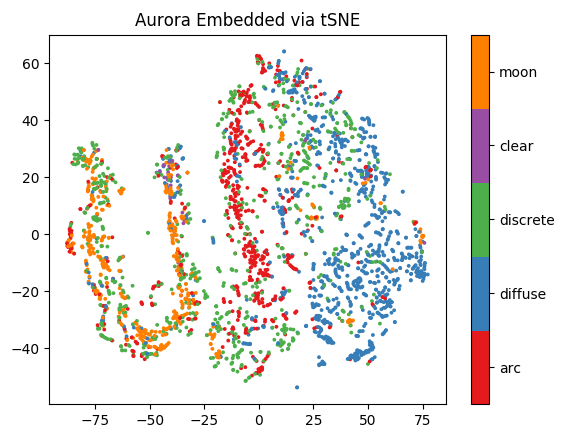} \\
			\small (a)
		\end{tabular} \qquad
		\begin{tabular}[b]{c}
			\includegraphics[width=.42\linewidth]{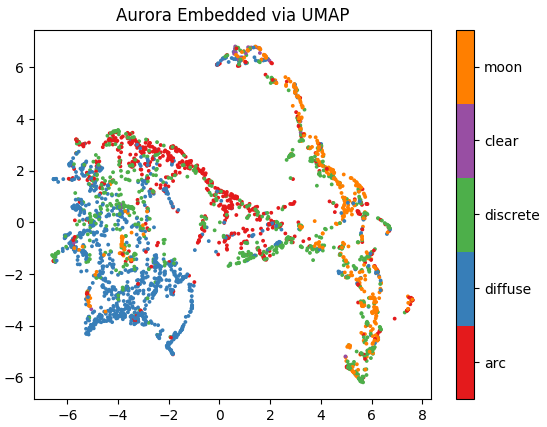} \\
			\small (b)
		\end{tabular}
		\caption{Latent space colored by image classes projected by (a) t-SNE (b) UMAP \label{tSNE}}
	\end{figure}

	We train the Res-AE on 2 months of THEMIS images from FSIM to encode our image class labelled dataset of 7700 images (we exclude images labelled as \emph{cloudy}, as these do not contain physical information on the ionosphere). Figure \ref{tSNE} shows the embedded latent representations for both the t-SNE and UMAP projections, colored by the image class labels.
	Both t-SNE and UMAP generate similar clusterings of the labelled image classes from the latent representations. They both most clearly cluster images identified as \emph{moon}, \emph{arc}, and \emph{diffuse}, with subsets of the \emph{discrete} images more closely associated with each of these clusters.
	
\begin{figure}
		\centering
		\begin{tabular}[b]{c}
			\includegraphics[width=.4\linewidth]{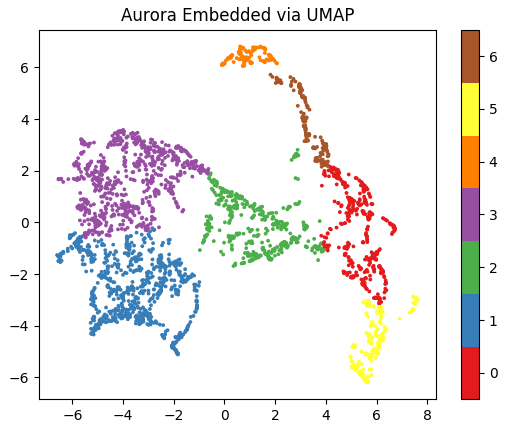} \\
			\small (a)
		\end{tabular} \qquad
		\begin{tabular}[b]{c}
			\includegraphics[width=.45\linewidth]{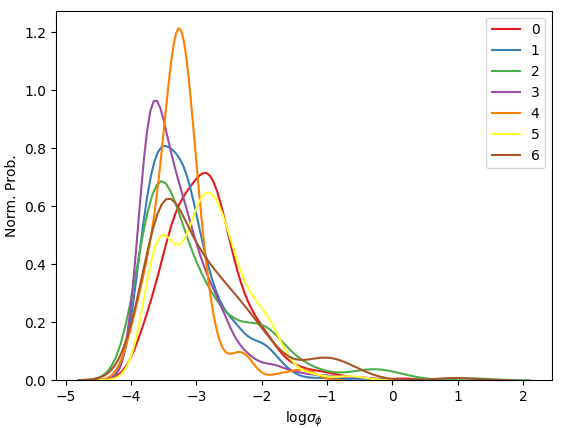} \\
			\small (b)
		\end{tabular}
		\caption{UMAP projection showing (a) spectral clustering (b) the corresponding log-normal distributions of $\sigma_{\phi}$ associated with each cluster. Colors and numbers in (b) correspond to the clusters identified in (a) by the spectral clustering algorithm applied to the UMAP projection. \label{sigmaphi}}
	\end{figure}

	We investigate the relation of the unsupervised image clusters to physically-relevant quantities related to the ionospheric electron density, namely the phase scintillation index ($\sigma_{\phi}$). Previous work has demonstrated that the intensity in the white light ASI images is correlated with the observed auroral precipitated energy  \cite[]{mende2009}, suggesting that ASI images can provide meaningful insight into physical processes occurring in the ionosphere, such as the localised fluctuations in electron density contributing to GNSS scintillations. Using spectral clustering, we identify distinct clusters in the projected latent space and look at the log-normal distribution of $\sigma_{\phi}$ measured by GNSS receivers co-located at the same site as the THEMIS ASI imagers. Figure \ref{sigmaphi}a demonstrates this approach on the UMAP projection.  Specific clusters of auroral images (e.g. clusters 0 and 5 in Figure \ref{sigmaphi}b) are correlated with significantly higher phase scintillation indices. These clusters correspond to a subset of the aurora classified as \emph{discrete} and \emph{arc} of Figure \ref{tSNE}b. Both of these clusters also contain images classified as \emph{moon}, suggesting that the brightness of auroral features (and thus the magnitude of the auroral precipitated energy) may be correlated with the magnitude of the $\sigma_{\phi}$. This analysis suggests that clustering in the low dimensional projection of the latent space can provide physically meaningful correlations with relevant physical parameters, such as $\sigma_{\phi}$.
	
	\section{Discussion and Outlook}
	While previous work has demonstrated that supervised deep learning approaches provide a useful method for analysing THEMIS images, here we demonstrate that unsupervised clustering techniques using a general-purpose Autoencoder may be a useful alternative approach. These results indicate that non-linear dimensionality-reduction techniques such as UMAP and t-SNE can provide meaningful lower-dimensional projections of the latent representation of aurora images that correlate with clusters associated with both human annotated image classes and physically meaningful parameters related to the ionospheric electron density variations. These results also indicate that specific dynamic structures in the aurora (as observed by the ASI) are more likely to correlate with GNSS phase scintillations. Such an approach seems to be site specific however, as we observed greater separation between images measured at different sites (RANK, FSIM) versus separation between image classes at a single site; additional masking of image edges may extend the applicability of the method to multiple sites. Future work will focus on applying this method to a larger image data set, further exploration of unsupervised auroral feature extraction from the t-SNE and UMAP projections correlated with high $\sigma_{\phi}$ values, and investigation of the correlation of clusters with other ionosopheric parameters, such as the total electron content. 
	
	\bibliographystyle{abbrvnat}
	\bibliography{neuripsresae}

\begin{thebibliography}{15}
\providecommand{\natexlab}[1]{#1}
\providecommand{\url}[1]{\texttt{#1}}
\expandafter\ifx\csname urlstyle\endcsname\relax
  \providecommand{\doi}[1]{doi: #1}\else
  \providecommand{\doi}{doi: \begingroup \urlstyle{rm}\Url}\fi

\bibitem[Aarons et~al.(2000)Aarons, Lin, Mendillo, Liou, and
  Codrescu]{aarons2000}
J.~Aarons, B.~Lin, M.~Mendillo, K.~Liou, and M.~Codrescu.
\newblock Global positioning system phase fluctuations and ultraviolet images
  from the {P}olar satellite.
\newblock \emph{Journal of Geophysical Research: Space Physics}, 105\penalty0
  (A3):\penalty0 5201--5213, 2000.

\bibitem[Clausen and Nickisch(2018)]{clausen2018}
L.~B. Clausen and H.~Nickisch.
\newblock Automatic classification of auroral images from the {O}slo {A}uroral
  {THEMIS} ({OATH}) data set using machine learning.
\newblock \emph{Journal of Geophysical Research: Space Physics}, 123\penalty0
  (7):\penalty0 5640--5647, 2018.

\bibitem[He et~al.(2016)He, Zhang, Ren, and Sun]{he2016}
K.~He, X.~Zhang, S.~Ren, and J.~Sun.
\newblock Deep residual learning for image recognition.
\newblock In \emph{Proceedings of the {IEEE} conference on computer vision and
  pattern recognition}, pages 770--778, 2016.

\bibitem[Hou and Vlontzos(2019)]{Vlontzos2019}
B.~Hou and A.~Vlontzos.
\newblock {R}es{N}et{AE}-https://github.com/farrell236/resnetae.
\newblock 2019.
\newblock URL \url{https://github.com/farrell236/ResNetAE}.

\bibitem[Jin et~al.(2014)Jin, Moen, and Miloch]{jin2014}
Y.~Jin, J.~I. Moen, and W.~J. Miloch.
\newblock {GPS} scintillation effects associated with polar cap patches and
  substorm auroral activity: {D}irect comparison.
\newblock \emph{Journal of Space Weather and Space Climate}, 4:\penalty0 A23,
  2014.

\bibitem[Jin et~al.(2016)Jin, Moen, Miloch, Clausen, and Oksavik]{jin2016}
Y.~Jin, J.~I. Moen, W.~J. Miloch, L.~B. Clausen, and K.~Oksavik.
\newblock Statistical study of the {GNSS} phase scintillation associated with
  two types of auroral blobs.
\newblock \emph{Journal of Geophysical Research: Space Physics}, 121\penalty0
  (5):\penalty0 4679--4697, 2016.

\bibitem[Maaten and Hinton(2008)]{maaten2008}
L.~v.~d. Maaten and G.~Hinton.
\newblock Visualizing data using t-{SNE}.
\newblock \emph{Journal of machine learning research}, 9\penalty0
  (Nov):\penalty0 2579--2605, 2008.

\bibitem[McInnes et~al.(2018)McInnes, Healy, and Melville]{mcinnes2018}
L.~McInnes, J.~Healy, and J.~Melville.
\newblock Umap: Uniform manifold approximation and projection for dimension
  reduction.
\newblock \emph{arXiv preprint arXiv:1802.03426}, 2018.

\bibitem[Mende et~al.(2009)Mende, Harris, Frey, Angelopoulos, Russell, Donovan,
  Jackel, Greffen, and Peticolas]{mende2009}
S.~Mende, S.~Harris, H.~Frey, V.~Angelopoulos, C.~Russell, E.~Donovan,
  B.~Jackel, M.~Greffen, and L.~Peticolas.
\newblock The{THEMIS} array of ground-based observatories for the study of
  auroral substorms.
\newblock In \emph{The THEMIS Mission}, pages 357--387. Springer, 2009.

\bibitem[Oksavik et~al.(2015)Oksavik, van~der Meeren, Lorentzen, Baddeley, and
  Moen]{oksavik2015}
K.~Oksavik, C.~van~der Meeren, D.~A. Lorentzen, L.~Baddeley, and J.~Moen.
\newblock Scintillation and loss of signal lock from poleward moving auroral
  forms in the cusp ionosphere.
\newblock \emph{Journal of Geophysical Research: Space Physics}, 120\penalty0
  (10):\penalty0 9161--9175, 2015.

\bibitem[Ronneberger et~al.(2015)Ronneberger, Fischer, and
  Brox]{Ronneberger2015}
O.~Ronneberger, P.~Fischer, and T.~Brox.
\newblock {U}-net: {C}onvolutional networks for biomedical image segmentation.
\newblock In \emph{International Conference on Medical image computing and
  computer-assisted intervention}, pages 234--241. Springer, 2015.

\bibitem[Spogli et~al.(2009)Spogli, Alfonsi, De~Franceschi, Romano, Aquino,
  Dodson, et~al.]{spogli2009}
L.~Spogli, L.~Alfonsi, G.~De~Franceschi, V.~Romano, M.~Aquino, A.~Dodson,
  et~al.
\newblock Climatology of {GPS} ionospheric scintillations over high and
  mid-latitude {E}uropean regions.
\newblock In \emph{Annales Geophysicae}. EGU, 2009.

\bibitem[van~der Meeren et~al.(2015)van~der Meeren, Oksavik, Lorentzen,
  Rietveld, and Clausen]{van2015}
C.~van~der Meeren, K.~Oksavik, D.~A. Lorentzen, M.~T. Rietveld, and L.~B.
  Clausen.
\newblock Severe and localized {GNSS} scintillation at the poleward edge of the
  nightside auroral oval during intense substorm aurora.
\newblock \emph{Journal of Geophysical Research: Space Physics}, 120\penalty0
  (12):\penalty0 10--607, 2015.

\bibitem[Yang et~al.(2019{\natexlab{a}})Yang, Tao, Han, and Liang]{qyang2019}
Q.~Yang, D.~Tao, D.~Han, and J.~Liang.
\newblock Extracting auroral key local structures from all-sky auroral images
  by artificial intelligence technique.
\newblock \emph{Journal of Geophysical Research: Space Physics},
  2019{\natexlab{a}}.

\bibitem[Yang et~al.(2019{\natexlab{b}})Yang, Wang, Song, and Gao]{xyang2019}
X.~Yang, N.~Wang, B.~Song, and X.~Gao.
\newblock {B}o{SR}: A {CNN}-based aurora image retrieval method.
\newblock \emph{Neural Networks}, 116:\penalty0 188--197, 2019{\natexlab{b}}.

\end{thebibliography}
\end{document}